\documentclass{elsart}
\usepackage{graphicx} 
\begin{document}
\begin{frontmatter} 

\title{Height and roughness distributions in thin films with Kardar-Parisi-Zhang
scaling}
\author[UFRJ]{Thereza Paiva}  
\ead{tclp@if.ufrj.br}
\address[UFRJ]{Instituto de F\'\i sica, Universidade Federal do Rio de Janeiro
\\
Caixa Postal 68528, 21941-972, Rio de Janeiro, RJ, Brazil}
\author[UFF]{F. D. A. Aar\~ao Reis\corauthref{cor1} }
\ead{reis@if.uff.br}           
\corauth[cor1]{Fax number: (55) 21-2629-5887 }
\address[UFF]{Instituto de F\'\i sica, Universidade Federal Fluminense,\\
Avenida Litor\^anea s/n, 24210-340, Niter\'oi, RJ, Brazil}
\date{\today}
\maketitle

\begin{abstract}
We study height and roughness distributions of films grown with discrete
Kardar-Parisi-Zhang (KPZ) models in a small time regime ($t/L^z \ll 1$) which is
expected to
parallel the typical experimental conditions. Those distributions are measured
with
square windows of sizes $8\leq r\leq 128$ gliding through a much
larger surface. Results for models with weak finite-size corrections indicate
that the absolute value of the skewness and the value of the kurtosis
of height distributions converge to $0.2\leq |S|\leq 0.3$ and $0\leq Q\leq 0.5$,
respectively. Despite the low accuracy of these results, they
give additional support to a recent claim of KPZ scaling in oligomer films.
However, there are significant finite-size effects in the scaled height
distributions of models with large local slopes, such as ballistic deposition,
which suggests that comparison of height distributions must not be used to rule
out KPZ scaling.
On the other hand, roughness distributions of the same models show good data
collapse, with negligible dependence on time and window size. The estimates of
skewness
and kurtosis for roughness distributions are $1.7\leq S\leq 2$ and $3\leq Q\leq
7$. A stretched exponential
tail was found, which seems to be a particular feature of KPZ systems in $2+1$
dimensions. Moreover, the KPZ roughness distributions cannot be fitted by those
of $1/f^{\alpha}$ noise. This study suggests that the roughness distribution is
the best option to test KPZ scaling in the growth regime, and provides
quantitative data for future comparison with other models or experiments.
\end{abstract}
 
\begin{keyword}
computer simulations; models of surface kinetics; surface roughness; height
distributions; roughness distributions; Kardar-Parisi-Zhang scaling
\PACS 68.35.Ct \sep 68.55.-a \sep 81.15.Aa \sep 05.50.+q 
\end{keyword}  

\end{frontmatter}
                   
\section{Introduction}
\label{intro}

In order to understand the basic mechanisms of thin films or multilayer growth,
it is helpful to compare the scaling properties of their surfaces with
those of statistical growth models, which are able to represent such features
without considering all details of the microscopic interactions
\cite{barabasi,krug,halpinhealy}. The usual starting point for this comparison
is to measure
the surface roughness and the associated scaling exponents. The
local roughness $w(r,t)$ is defined as the rms fluctuation of the interface
height averaged over windows of linear size $r$ spanning the surface at time
$t$. For fixed $t$, it scales as
$w \sim r^{\alpha}$ for $r\ll L$, where $\alpha$ is the roughness
exponent and $L$ is the total surface length. On the other hand, the global
roughness $\xi (t)$ is defined as the
rms height fluctuation over the whole surface [$w(r,t)$ for
$r\to L$], and scales as $\xi \sim t^{\beta}$, where $\beta$ is called the
growth exponent. Instead of the local roughness $w$, one may use height-height
correlation functions, which have similar scaling properties.

A large number of experimental works have already obtained scaling exponents
consistent with those of the Kardar-Parisi-Zhang (KPZ) theory
\cite{kpz} - see e. g. Refs.
\protect\cite{krim,wang,miettinen,paniago,tsamouras,palasantzas,marcilei,ebothe}.
However, the available scaling region of experimental data is usually very
small, including less than one order of magnitude of length or time. 
In some cases, a crossover from another dynamics to KPZ is suggested
\cite{kleinke,marta,nara,lita}. The difficulties to find accurate scaling laws
for the local roughness are also present in the study of growth
models \cite{chamereis}, thus the main limitation is the method and not the
nature (experimental or theoretical) of the data.
At this point, the methods that provide accurate estimates of exponents for
theoretical models are not helpful because they usually consider steady
state data, where film thicknesses are very large and the full system
is highly correlated ($t/L^z\gg 1$, where $z=\alpha /\beta$ is the dynamic
exponent). However, this regime is not typical of experiments, where the
thickness and the lateral correlation length are much smaller than the lateral
size of the deposit ($t/L^z\ll 1$).

This scenario motivates the proposal of alternative or complementary
approaches to compare results of theoretical models and experimental data in the
growth regime. Two alternatives are suggested by recent theoretical work: the
study of the full height distributions \cite{shim,marinari,kpz2d}, and the
study of the full roughness distributions
\cite{foltin,racz,antal96,antalpre,marinari2,distr1}. For instance, steady state
roughness distributions of Gaussian
interfaces were already helpful in the analysis of some controversial systems
\cite{bramwell,rosso,sigma} and in the fit of experimental distributions of
depinning interfaces \cite{moulinet}. Moreover, height distributions for
KPZ systems in $1+1$ dimensions \cite{prahofer} were observed to fit those
of combustion fronts propagating in sheets of paper \cite{miettinen}, in the
growth and in the steady state regimes. However, as far as we know, for
three-dimensional ($2+1$-dimensional) KPZ systems, the currently available
numerical results are limited to steady state distributions
\cite{marinari,kpz2d,distr1}. On the other hand, experience in $1+1$
dimensions shows that roughness distributions of EW interfaces in the growth
regime are very different from the steady state ones \cite{antal96}, and that
KPZ scaling is also very different in these regimes \cite{kimmoorebray,krug92}.

In this work, we will analyze numerical data of height and roughness
distributions of discrete KPZ models in the growth regime ($t/L^z\ll 1$) in
$2+1$ dimensions.
Our aims are to test the reliability of those distributions to determine the
KPZ universality class and to provide a set of quantitative data for comparison
with other systems. In order that these results can be compared with
experimental data, we will calculate distributions in narrow windows gliding
through a much larger substrate. In other words, we will adopt the so-called
window boundary conditions (WBC) \cite{antalpre}.

First we will show discrepancies in the scaled height distributions of three
discrete KPZ models, although the estimates of the absolute value of the
skewness and of the kurtosis are not very different when two of them are
compared. Despite having a large error bar, the estimate of the skewness for
these models agrees with that of  Refs. \protect\cite{tsamouras,palasantzas}
for oligomer films. However, the data for the third model (ballistic deposition)
exhibit strong finite-size corrections, which suggests that the height
distribution cannot be used to rule out KPZ growth.
On the other hand, we will show that roughness distributions in WBC show weak
finite-size and finite-time corrections, providing a good data
collapse for all KPZ models analyzed here. Estimates of the
skewness and kurtosis of the universal distribution in the growth regime will
be provided here, which we expect to be helpful for future identification of the
KPZ universality class in experiments.

The rest of this work is organized as follows.
In Sec. 2, we will present the theoretical models considered here and details of
the simulation procedure. In Sec. 3, we will present the results for the
heights distributions. In Sec. 4, we will present the results for the roughness
distributions. In Sec. 5 we summarize our results and present our conclusions. 

\section{The KPZ equation and the lattice models}
\label{models}

The Kardar-Parisi-Zhang equation \cite{kpz}
\begin{equation}
{{\partial h}\over{\partial t}} = \nu{\nabla}^2 h + {\lambda\over 2}
{\left( \nabla h\right) }^2 + \eta (\vec{x},t) ,
\label{kpz}
\end{equation}
was proposed as a hydrodynamic description of processes where the local slope
of the surface has significant effects on the growth velocity. In Eq.
(\ref{kpz}), $\nu$ is a surface tension, $\lambda$ represents the excess
velocity (due e. g. to lateral growth) and $\eta$ is a Gaussian white noise.
For growth in two-dimensional substrates, no exact solution is currently
available, and the best known estimates of KPZ exponents, $\alpha\approx 0.39$
and $\beta\approx 0.23$, were obtained numerically \cite{marinari,kpz2d}.

Several stochastic lattice models, which are designed to represent
microscopic aggregation processes, belong to the KPZ class, i. e. in the limit
of large lengths and long times they are described by the KPZ 
equation. In this work, we will consider four models in this class: two
restricted solid-on-solid (RSOS) models \cite{kk}, the etching model of Mello
et al \cite{mello} and ballistic deposition (BD).
In the RSOS model, the incident particle may stick at the top of the column of
incidence if the differences of heights between neighboring columns do not
exceed a certain value ${\Delta h}_{MAX}$. Otherwise, the aggregation attempt is
rejected. Here, we will consider the cases ${\Delta h}_{MAX}=1$ and ${\Delta
h}_{MAX}=2$, hereafter called RSOS and RSOS2, respectively. 
The etching model of Mello et al \cite{mello} is considered in its growth
version: at each growth attempt, a column $(x,y)$, with
current height $h(x,y)\equiv h_0$, is randomly chosen; subsequently, its height
is increased by one unit ($h(x,y)\rightarrow h_0+1$), and any neighboring
column whose height is smaller than $h_0$ grows until its height becomes
$h_0$. Finally, in BD, each incident particle is released from a randomly chosen
position far above the substrate, follows a trajectory perpendicular to the
surface and sticks upon first contact with a nearest neighbor occupied
site \cite{fv,vold}.

In all models, one time unit corresponds to $L^2$ random
selections of columns of incidence in a lattice of linear size $L$,
independently of the average height increase. Simulations were done in 
lattices with $L=1024$, up to times $t_{max}=8\times {10}^3$ for the RSOS, RSOS2 
and BD models and $t_{max}=6\times {10}^3$ for the etching model. Nearly
$100$ different deposits were generated for each model, starting from flat
substrates. Our previous experience with simulation of those models
\cite{kpz2d,distr1} show that these times are sufficiently small compared to
the relaxation times to the steady states, so that the deposits are certainly
in the growth regime. Periodic boundary conditions were considered, but they
are not expected to affect our conclusions because we have analyzed results from
windows sizes much smaller than $L$ and the lateral correlation lengths are also
much smaller than $L$.

At time intervals typically $\Delta t=1000$, the local roughness and height
distributions were collected for various window sizes $r$, with square windows.
To be more precise, we calculated $P_h(\Delta h)$ and $P_w(w_2)$, so that
$P_h(\Delta h)dh$ is the probability that the height relative to the window
average
($\Delta h=h-\overline{h}$) is in the interval $[\Delta h,\Delta h+dh]$, and
$P_w(w_2)dw_2$ is the probability that the squared roughness $w_2$ is in the
interval $[w_2,w_2+dw_2]$.   
For each $r$, the center of the window was allowed to occupy all columns of the
substrate, and height fluctuations inside that window were measured.
For small $r$, a large number of different microscopic environments is present
in a large lattice and, consequently, accurate results are obtained with a
small number of realizations of the growth process. 

\section{Height distributions}

In the analysis of the height distributions of KPZ models, one should care about
the consequences of the symmetry of the KPZ equation (\ref{kpz}) under the
transformation $h\to -h$, $\lambda\to -\lambda$ \cite{kpz2d}. While RSOS
models are represented by KPZ equations with negative $\lambda$, BD and the
etching  model have positive $\lambda$ \cite{kpz2d,krug92}. Thus the
transformation $h\to -h$ should be applied to one of these sets of models for
the comparison of their height distributions.

In Fig. 1 we show a log-linear plot of the height distributions for the RSOS,
the etching and the BD models at $t=4000$ and window size $r=128$. $P_h(\Delta
h)$ is shown for models with $\lambda >0$ (etching and BD) and $P_h(-\Delta h)$
is shown for the model with $\lambda <0$ (RSOS). We do not observe a very good
data collapse between the data for these models, but only a reasonable
agreement near the peaks. Deviations of the BD distribution are particularly
large in its tails. Discrepancies are found in other window sizes, and they are
larger for BD.
\begin{figure}[!h]
\includegraphics[width=15cm]{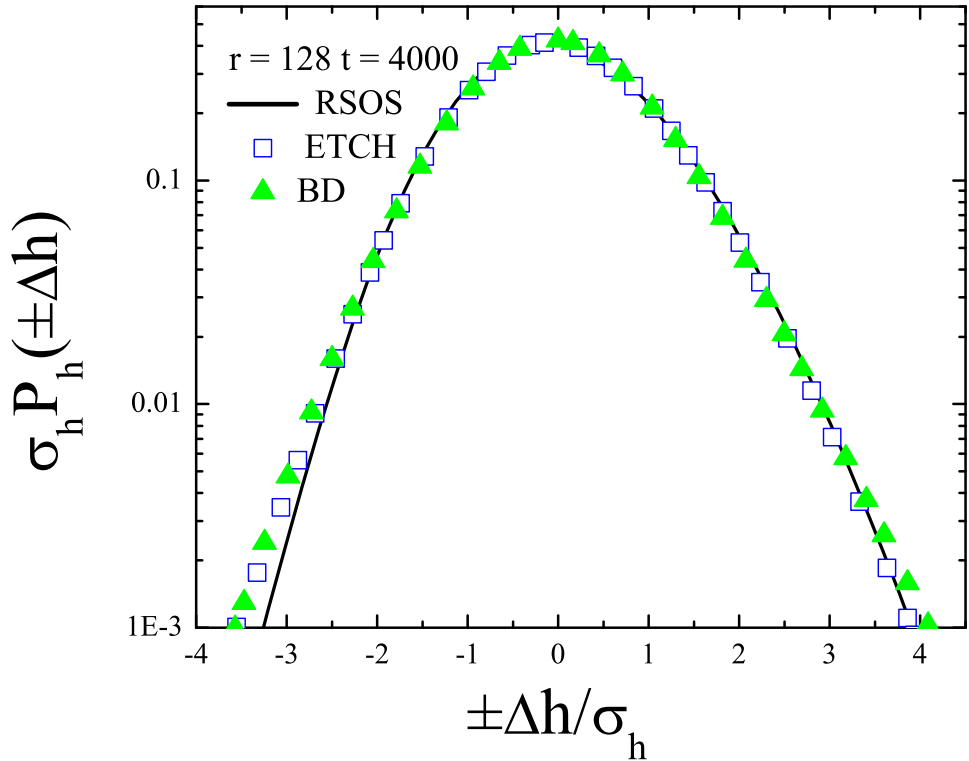}
\caption{Normalized height distributions
$\sigma_h P_h(\pm \Delta h)$ as a function of $\pm {\Delta h}/\sigma_h$ 
for
$r=128$ and $t=4000$, where  $\Delta h\equiv h- \left< h \right>$ and 
$\sigma_h$ is the rms fluctuation of $\Delta h$. Plus signs
for etching model (empty squares) and BD (full triangles), minus signs for 
RSOS model (solid curve).
}
\label{fig1}
\end{figure}

A quantitative characterization of each distribution can be done by estimating
the skewness
\begin{equation}
S \equiv \frac{\langle {\left( \Delta h\right)}^3 \rangle}
{{\langle {\left( \Delta h\right)}^2 \rangle}^{3/2}}
\label{skewh}
\end{equation}
and kurtosis
\begin{equation}
Q \equiv \frac{\langle {\left( \Delta h\right)}^4 \rangle}
{{\langle {\left( \Delta h\right)}^2 \rangle}^2} - 3 .
\label{curth}
\end{equation}
The former is a measure of the asymmetry of the distribution, while the latter
is a measure of the weight of its tails when compared to a Gaussian.

In Fig. 2a we plot $\pm S$ for those models ($+S$ for models with $\lambda>0$
and $-S$ for models with $\lambda<0$), obtained in $t=4000$. The values for the
RSOS and the etching models do not show a significant size dependence for $1\ll
r\ll L$ (simulations in $L=1024$). Extrapolation to large $r$ ($1/r\to\infty$)
suggests $|S| \approx 0.2$. However, there is a large difference between those
models and
BD, whose skewness has a remarkable dependence on the window size. Such
behavior was also observed in steady state data for BD \cite{kpz2d}, thus this
discrepancy must not be viewed as a failure of universality, but an expected
feature for a model with typically large corrections to scaling.
\begin{figure}[!h]
\includegraphics[width=15cm]{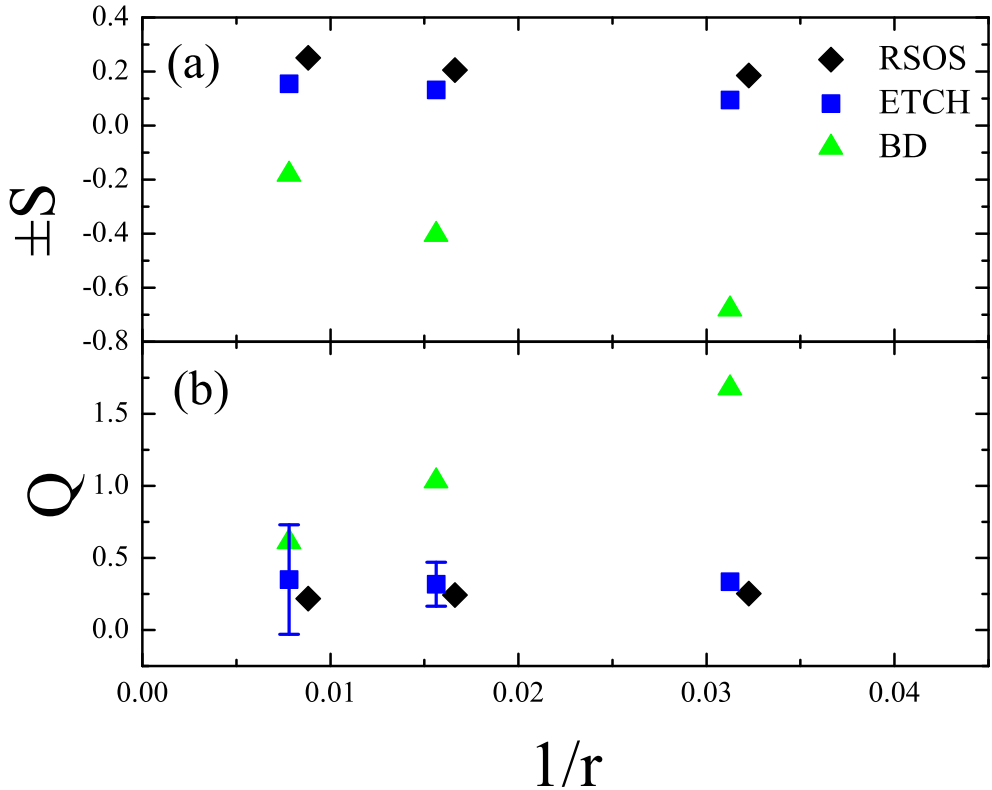}
\caption{Skewness $\pm S$ (a) and kurtosis $Q$ (b) of height distributions as a
function of inverse window size at $t=4000$ for
RSOS (diamonds), etching (squares) and ballistic deposition (triangles). Minus  
sign of $S$ only for the RSOS model, whose data have been shifted horizontally
by $0.001$ for clarity. Error bars are of the same size of the data points,
except where indicated.
}
\label{fig2}
\end{figure}

In Fig. 2b, we show $Q$ for those models as a function of inverse window size.
Fluctuations are larger, thus it is difficult to obtain a very accurate
asymptotic estimate. Anyway, it is clear from Fig. 2b that $0<Q<0.5$ for these
KPZ systems, which may be eventually used to test the possibility of this
universality class in other models or experiments.
The results obtained for other times ($t=1000$ to $t=6000$) show the same
trends of those in Figs. 2a and 2b.

Tsamouras et al \cite{palasantzas} obtained $0.2\leq S\leq 0.5$ for
the height distributions of oligomer films, and $\alpha\approx 0.45$ from the
scaling of local roughness. This exponent is near the KPZ value, and the
asymmetry of the height distribution was considered as further evidence of KPZ
scaling, since no theoretical value of $S$ was known at that time. We believe
that the claim of KPZ scaling is reinforced by our estimate of $S$, despite the
large error bars and the discrepancies in the values of $S$ for different
models. On the other hand, the same uncertainties and discrepancies show that
height distributions may not be considered reliable evidence against KPZ
scaling.

It is also interesting to compare the above results with those for KPZ models in
the steady states. There, $|S|=0.26\pm 0.01$ and $Q=0.134\pm 0.015$ were
obtained
\cite{marinari,kpz2d}. Our estimates have much larger error bars as an effect of
finite-time and finite-size corrections, which are much stronger than in the
steady states. Thus, we are not able to decide whether there is some difference
between the height distributions in the growth regime and in the steady state.
Anyway, we recall that distributions in these regimes are different in $1+1$
dimensions \cite{krug92}, thus the present analysis of the growth regime is
essential for a comparison with experimental data.

\section{Roughness distributions}

In Fig. 3 we show the scaled squared roughness distributions for the RSOS,
RSOS2, etching and BD models, obtained in window sizes $r=64$ and $t=4000$.
Here, $\langle w_2\rangle$ is
the average squared roughness and $\sigma_{w}\equiv \sqrt{ \left< {w_2}^2
\right> -
{\left< w_2\right>}^2 }$ is the rms deviation. In this case, a good data
collapse is observed in the peaks and in the tails of the distributions.
\begin{figure}[!h]
\includegraphics[width=15cm]{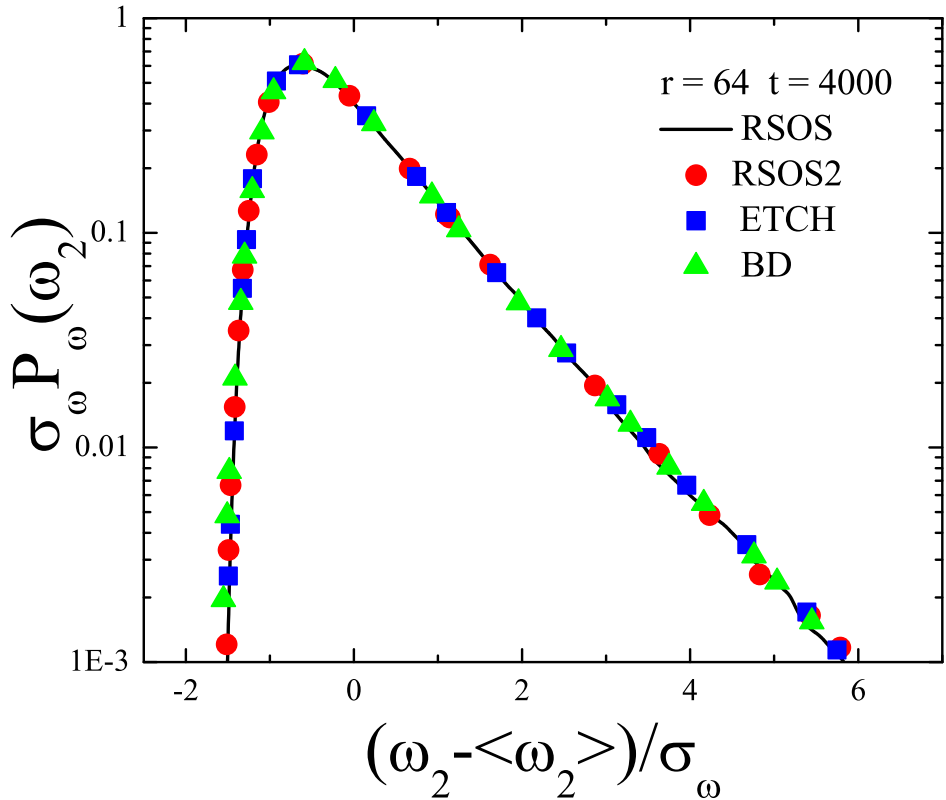}
\caption{Normalized squared roughness distribution $\sigma_w P_w
\left(w_2\right)$ as a function of $( w_2 - \left< w_2 \right> ) / \sigma_w$
for window size $r=64$ and $t=4000$, for RSOS (line),
RSOS2 (circles),
etching (squares) and ballistic deposition (triangles) models. }
\label{fig3}
\end{figure}

This is confirmed by the values of the skewness and the kurtosis of those
distributions, which are shown if Fig. 4a and 4b, respectively, as a function
of inverse window size. Although the size dependences of $S$ and $Q$ are
not negligible, the convergence of the results for different
KPZ models as $r\to\infty$ is suggested, which gives universal values of $S$
and $Q$. The trend of the data for all models suggest $1.7\leq S\leq 2.0$ and
$3 \leq Q\leq 7$.
\begin{figure}[!h]
\includegraphics[width=15cm]{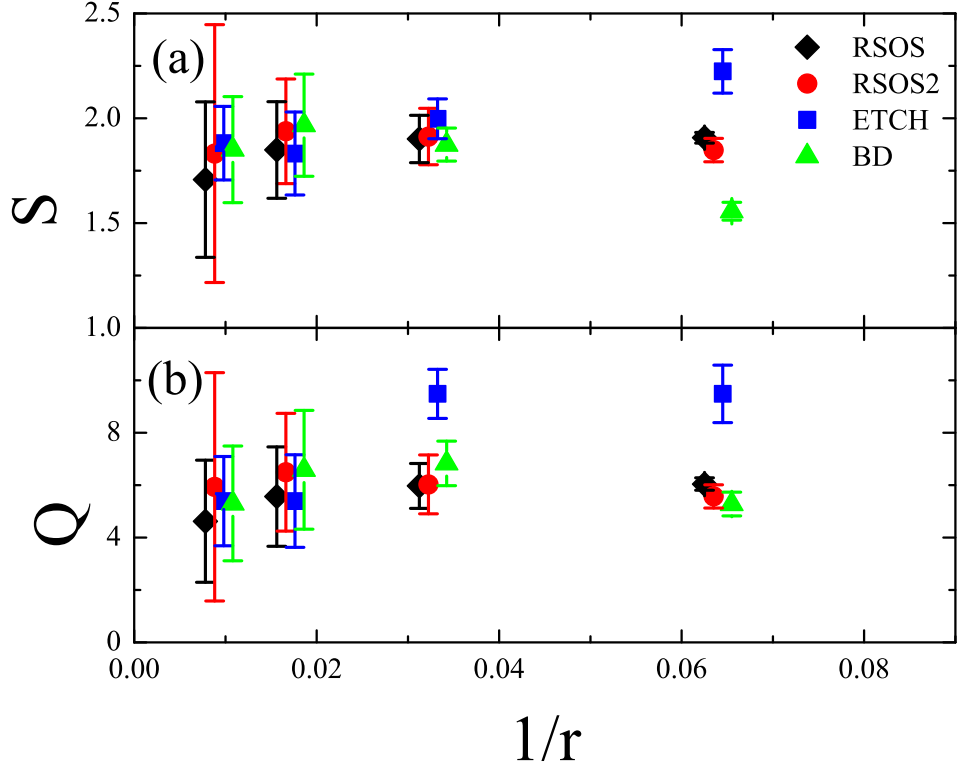}
\caption{
Skewness  (a) and kurtosis (b) of roughness distributions as a function of 
inverse window size at $t=4000$ for
RSOS (diamonds), RSOS2 (circles), etching (squares) and ballistic deposition  
(triangles). Data
for RSOS2, etching and ballistic deposition have been shifted horizontally
by $0.001$, $0.002$ and $0.003$ respectively for clarity.
}
\label{fig4}
\end{figure}

The universal KPZ distribution in the
growth regime is slightly different from that in the steady state. This is
illustrated in Fig. 5, where we show the normalized
distributions for the RSOS model and the BD model in the growth regime, with
$r=64$ and different times, and the distribution for 
the RSOS model in the steady state, for lattice size $L=256$. The plot of Fig. 5 
is also helpful to show that the dependence of the scaled roughness
distribution on time is very small. Although visual
inspection shows deviations in the tails of the distributions of different
regimes, the
above values of the skewness and kurtosis have large uncertainties and,
consequently, they are not able to exclude the steady state values $S= 1.70\pm
0.02$ and $Q=5.4\pm 0.3$ \cite{distr1}.
\begin{figure}[!h]
\includegraphics[width=15cm]{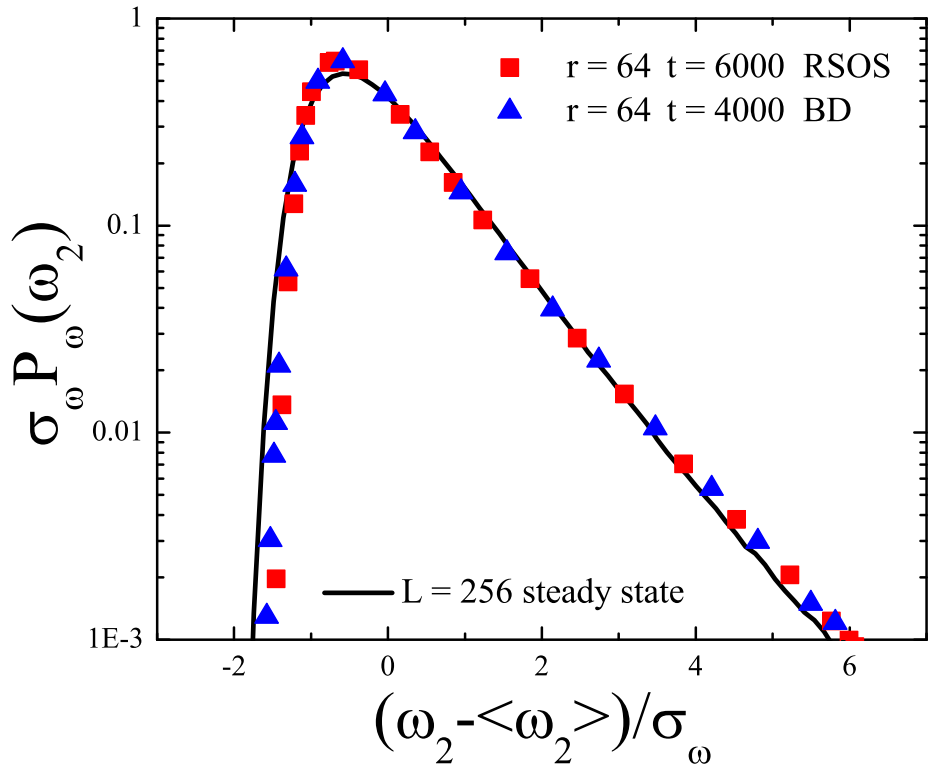}
\caption{
Normalized squared roughness distributions for the RSOS model in the steady
state with lattice size $L=256$ (solid curve), and for RSOS and BD in the   
growth regime with $r=64$ and different times ($t=4000$ for BD - triangles -
and $t=6000$ for
RSOS - squares).}
\label{fig5}
\end{figure}

Another important feature of the KPZ roughness distribution is the
stretched exponential tail, as suggested by a careful inspection of the
log-linear plots of Figs. 3 and 5. In order to find a reliable fit of this tail,
it is reasonable to
assume that the scaling function decay has the general form $P_w\left( x\right)
\sim \exp{\left( -Ax^\gamma \right)}$. Thus, at a given range of values of the
variable $x\equiv \left( w_2- \left< w_2 \right>\right) / \sigma_w$, the
estimate of the exponent $\gamma$ is given by
\begin{equation}
\gamma\left( x\right) = \frac{\ln{\left[
{ \ln{\left( \Psi\left( x\right) \right)} }/
{ \ln{\left(  \Psi\left( x-\Delta\right)
\right)} }
\right]}}
{\ln{\left[ x/\left( x-\Delta\right)\right]}} ,
\label{defgama}
\end{equation}
with constant $\Delta$, where
$\Psi(x)=\sigma_w P_w(x)$.
In Fig. 6 we show the effective exponents obtained from the data of the
RSOS and the etching models in $r=64$ as a function
of $1/x^2$. The trend for large $x$ suggests $0.85\leq \gamma\leq 0.9$.
\begin{figure}[!h]
\includegraphics[width=15cm]{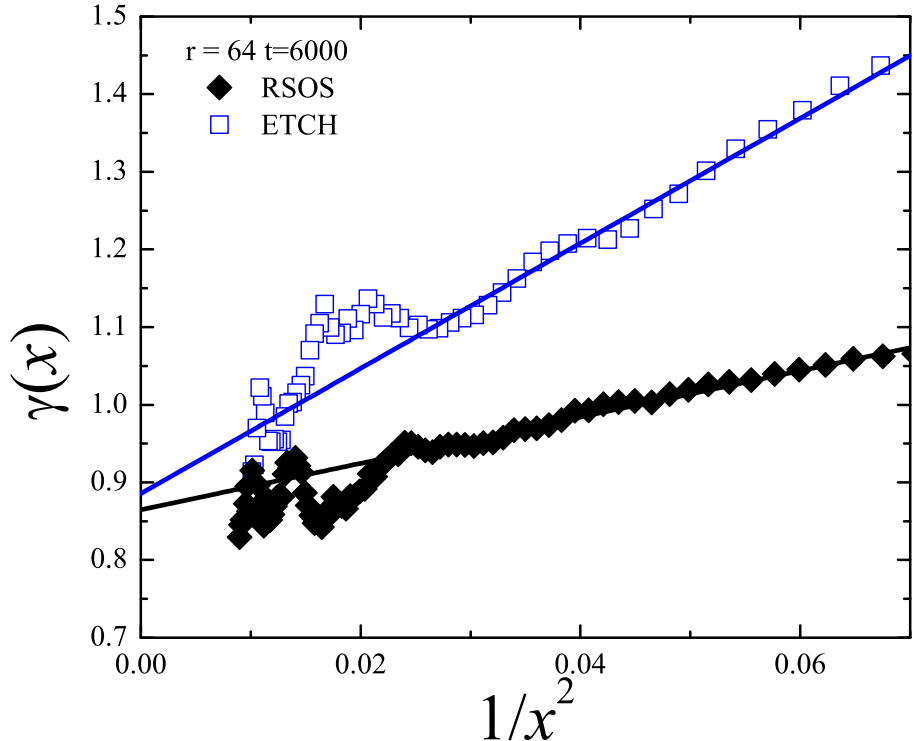}
\caption{
The effective exponent $\gamma\left( x\right)$ as a function of $1 / x^2$, where
$x=(w_2- \left< w_2
\right>) /
\sigma_w$. Data are
for $t=6000$ and $r=64$, full diamonds for the RSOS  and empty squares for the
etching model.}
\label{fig6}
\end{figure}

In order to appreciate the relevance of this result, we recall that other
systems which are represented by linear growth equations have roughness
distributions with simple exponential decays\cite{racz,distr1}, of the type
$\exp{\left( -Cx\right)}$ ($\gamma = 1$). This also occurs in
$1/f^\alpha$-noise interfaces \cite{antalpre}. For the particular case of EW
growth in $2+1$ dimensions, a sharply peaked roughness distribution is
expected\cite{racz}. These are examples of Gaussian interfaces. Moreover, a
simple exponential tail is also obtained in roughness distributions of models
in the class of the fourth order nonlinear growth equation in $2+1$ dimensions
\cite{distr1}.

The comparison which these widely studied systems suggests that the stretched
exponential tail in the roughness distribution may be a signature of KPZ
scaling in $2+1$ dimensions. It is also obtained in the steady state of KPZ
systems, as expected from universality grounds \cite{distr1}. Thus, if highly
accurate experimental data eventually shows a stretched exponential tail, it
may be viewed as strong evidence of KPZ scaling.

Unfortunately, current theoretical approaches could not to provide a clear and
consistent picture of the KPZ scaling in $2+1$ dimensions
\cite{lassig,colaiori,fogedby}. At some points, they lead to contradictory
results and show discrepancies with highly accurate numerical data
\cite{marinari,kpz2d,marinari2}. Consequently, we cannot provide a reliable
explanation for this stretched exponential tail, but only to state that it is
probably related to the non-Gaussian behavior of these KPZ interfaces
\cite{kpz2d,distr1}.

A final important point is that the KPZ roughness distribution in the growth
regime cannot be fitted by $1/f^{\alpha}$-noise distributions for any value of
$\alpha$, similarly to the steady state distribution. The stretched exponential
tail is a strong reason to expect that such fit is impossible. However, this
conclusion also follows from our high value for the skewness: in
$1/f^{\alpha}$-noise distributions in $2+1$ dimensions, the maximum possible
value of the skewness is $\sqrt{2}$ \cite{antalpre}, which is far below the
lower bound of our estimate, $S=1.7$. 

\section{Conclusion}
\label{conclusion}

We studied height and roughness distributions of discrete KPZ models in the
growth regimes, i. e. in the regimes where correlations along the substrate
directions are developed but the correlation length is still much smaller than
the substrate size.

Even considering models with typically weak scaling corrections, the height
distributions of those models show some deviations, thus we were
not able to extract very accurate universal values of amplitude ratios to
characterize them: the absolute value of the skewness is in the range
$[0.2,0.3]$ and the kurtosis is in the range $[0,0.5]$. The value of the
skewness agrees with that of oligomer films in Refs.
\protect\cite{tsamouras,palasantzas}, which provides additional support to
their claim of KPZ scaling. However, the data for ballistic deposition shows
large deviations from these values, which suggests that height distributions
may be misleading for a test of KPZ scaling.

On the other hand, roughness distributions of different models show a good data
collapse, and we estimate their skewness $1.7\leq S\leq 2.0$ and kurtosis
$3 \leq Q\leq 7$. A stretched exponential
tail is found, which seems to be a particular feature of KPZ systems in $2+1$
dimensions. We conclude that roughness distributions are much superior than
height distributions for tests of KPZ scaling.
We expect that these results can be useful for a reliable quantitative
comparison with available experimental data and that they motivate further
analysis of scaling properties of thin films and/or multilayers surfaces.

\vskip 1cm

{\bf Acknowledgements}

FDAAR thanks Prof. Zoltan R\'acz for helpful discussion and suggestions.

This work was partially supported by CNPq and FAPERJ (Brazilian agencies).


\end{document}